\documentclass{article}
\usepackage{graphicx}
\usepackage{cite}
\usepackage{epsfig}
\usepackage{amsfonts}
 \usepackage{amssymb}
\usepackage{amsmath}
\usepackage{slashed}

\usepackage{hyperref}
\usepackage{dcolumn}        
\usepackage{bm}            		 
\usepackage{amstext}
\usepackage[dvipsnames]{xcolor}
\usepackage{color} 
\usepackage{transparent}
\usepackage{rotating}
\usepackage{array}
\usepackage[T1]{fontenc}
\usepackage[utf8]{inputenc}
\newcommand\mybar{\kern1pt\rule[-\dp\strutbox]{.8pt}{\baselineskip}\kern1pt}
\usepackage{float}

\newcolumntype{P}[1]{>{\centering\arraybackslash}p{#1}}

\makeatletter
\newcommand{\vast}{\bBigg@{4}}
\newcommand{\Vast}{\bBigg@{5}}
\makeatother

\date{\today}

\begin{document}
\begin{center}

{\Large \bf Deconstructing scaling virial identities in General Relativity: \\ spherical symmetry and beyond}
\vspace{0.8cm}
\\
{Carlos A. R. Herdeiro$^{1}$, Jo\~ao M. S. Oliveira$^{2}$, 
Alexandre M. Pombo$^{1,3}$, Eugen Radu$^{1}$,  \\
\vspace{0.3cm}
$^{1}${\small Departamento de Matem\'atica da Universidade de Aveiro and } \\ {\small  Centre for Research and Development  in Mathematics and Applications (CIDMA),} \\ {\small    Campus de Santiago, 3810-183 Aveiro, Portugal}
\\
\vspace{0.3cm}
$^{2}${\small Centro de Matem\'atica, Universidade do Minho, 4710-057 Braga, Portugal}
\\
\vspace{0.3cm}
$^{3}${\small Departamento de F\'\i sica, Instituto Superior T\'ecnico - IST,} \\ {\small Universidade de Lisboa - UL,
Avenida Rovisco Pais 1, 1049-001 Lisboa, Portugal}
}
\end{center}

	\begin{abstract}   
Derrick-type virial identities, obtained via dilatation (scaling) arguments, have a variety of applications in field theories. We deconstruct such virial identities in relativistic gravity showing how they can be recast as self-evident integrals of appropriate combinations of the equations of motion. In spherical symmetry, the appropriate combination and gauge choice guarantee the geometric part can be integrated out to  yield a master form of the virial identity as a non-trivial energy-momentum balance condition, valid for both asymptotically flat black holes and self-gravitating solitons, for any matter model. Specifying the matter model we recover previous results obtained via the scaling procedure. We then discuss the more general case of  stationary, axi-symmetric, asymptotically flat black hole or solitonic solutions in General Relativity, for which a master form for their virial identity is proposed, in a specific gauge but regardless of the matter content.  In the flat spacetime limit, the master virial identity for both the spherical and axial cases reduces to a balance condition for the principal pressures, discussed by Deser.
	\end{abstract}

\medskip

\tableofcontents

\newpage

%
%
\section{Introduction}\label{S1}

A far reaching question in any field theory is the possible existence of particle-like solutions, or \textit{solitons}. Generic arguments addressing this question are therefore desirable. One such insightful argument was put forward by Deser~\cite{Deser:1976wq}. For a Minkowski spacetime classical field theory, with Cartesian spatial stress-energy tensor $T_{ij}$, $i,j=1,2,3$, any localised, finite energy, everywhere regular solution obeys\footnote{See also the pioneering work of Von Laue~\cite{Laue:1911lrk}.}
\begin{equation}
\int_{\mathbb{R}^3} T_i^i \, d^3x =0 \ .
\label{Deser}
\end{equation}
This simple relation means that the sum of the principal pressures integrated over the whole space vanishes. Thus, if a soliton exists, the corresponding field(s) must be under compression in some regions and under tension in  other regions~\cite{Gibbons:1990um}. In other words, eq.~\eqref{Deser} is a \textit{pressure balance} condition.

\medskip

An apparently different generic argument establishing a condition for solitons in (scalar) field theories was provided by Derrick~\cite{derrick1964comments}. By assuming the existence of an hypothetical solitonic solution and \textit{spatially scaling it}, an integral condition for the field's spatial grandients and potential energy is obtained. As it turns out, this scaling (or virial~\cite{Herdeiro:2021teo}) identity precisely coincides with~\eqref{Deser}, as we shall illustrate below.

The connection between Deser's and Derrick's arguments is not accidental. Considering static configurations in a classical field theory with a conserved  energy momentum tensor ($\partial_i T^i_{\ j}=0$) one can define a dilatation or scaling current, $j^i=x^jT^i_{\ j}$~\cite{Forger:2003ut}, which is \textit{locally} conserved ($\partial_ij^i=0$) iff the energy-momentum tensor is  traceless ($T^i_{\ i}=0$). For a generic classical field theory with $T^i_{\ i}\neq0$, on the other hand, it is still true that the integral over a spacelike surface $\int \partial_ij^i d^3x$ vanishes for localised, static, finite energy, regular configurations, even though the integrand does not, implying~\eqref{Deser} -  $cf.$ Section~\ref{Section33} below. Thus, one may view~\eqref{Deser} as a \textit{global} conservation law under dilatations, providing a connection with Deser's argument. 

\medskip

Introducing gravity opens up new possibilities for solitons. Certain field theory models minimally coupled to Einstein's gravity allow for self-gravitating solitons. Moreover, relativistic gravity allows other localised configurations, namely black holes (BHs).  It is thus natural to ask whether~\eqref{Deser} can be generalized to General Relativity (GR). That is, do self-gravitating solitons, or BHs, in any field theory minimally coupled to Einstein's gravity obey a curved spacetime generalization of~\eqref{Deser}, $i.e.$ an \textit{energy-momentum} balance condition? If so, moreover, is it related to Derrick-type virial identities in relativistic gravity?

In this work we shall  propose a positive answer to both these questions, at least in some circumstances, in particular when appropriate gauge choices are made. We will generalize~\eqref{Deser} to GR models accommodating BHs or self-gravitating solitons. Concretely, we show that any static, spherical BH or soliton, under appropriate regularity conditions on the horizon/origin and asymptotically, describable in the metric gauge
\begin{eqnarray}
\label{metric-Schw}
 ds^2=-N(r)\sigma^2(r) dt^2+\frac{dr^2}{N(r)}
+r^2( d\theta^2+\sin^2\theta d\varphi^2  )\ , \qquad 
 \ \ \ N(r)\equiv 1-\frac{2m(r)}{r} \ ,
\end{eqnarray}
obeys, for \textit{any matter model}, the identity
\begin{eqnarray}
\int_{r_H}^\infty
r^2 \sigma
\left\{
\left(1-\frac{r_H}{r}\right)
\left[
2(T_t^t-T_\mu^\mu)+\left(\frac{1}{N}-1\right)(T_r^r-T_t^t)
\right]
-\frac{2r_H}{r}T_r^r
\right\}dr=0 \ ,
\label{tmnid1}
\end{eqnarray}
where $r_H$ is the event horizon radial coordinate ($r_H=0$) for solitons. We emphasise that the identity~\eqref{tmnid1} shall be derived without specifying any matter model. Then, considering specific matter models, we show it coincides with the virial identity derived by Derrick's scaling argument. This holds for all examples we have considered in~\cite{Herdeiro:2021teo}. Thus, we \textit{propose} that~\eqref{tmnid1} is both the Derrick-type virial identity of the model (under the appropriate boundary conditions) and the generalization of the Deser identity~\eqref{Deser}, to which~\eqref{tmnid1} reduces in the flat spacetime limit ($r_H=0$, $N=\sigma=1$), in the spherically symmetric guise of the former.

\medskip

We will also discuss the generalization of this analysis for a generic class of stationary, axially symmetric self-gravitating solitons and BHs in GR minimally coupled to matter models. Concretely, we show that asymptotically flat equilibrium solutions (solitons or BHs under appropriate regularity conditions), described by a stationary, axi-symmetric and circular line element (cf.~\eqref{AxmetricNum} below) obey,  for \textit{any matter model}, an identity which coincides with the one obtained by using Derrick's scaling argument for the specific models we have studied and reduces to Deser's identity~\eqref{Deser}  in the flat spacetime limit. In this case, however, we could not express the identity -- given by eq.~\eqref{f2} below -- solely as an energy-momentum balance condition. Rather, it is expressed as a self-evident identity, an integral over an appropriate combination of the Einstein equations.

\medskip

 Let us remark that there has been a number of previous works on virial identities in GR,  notably~\cite{Chandrasekhar:1965gcg,1973ApJ...182..335B,Gourgoulhon_1994,1979ApJ...227..307V} (see also~\cite{2013rrs..book.....F}), within the perspective of generalizing to the relativistic context the usual virial theorem of classical mechanics - see~\cite{1978vtsa.book.....C} for an overview. Our work focuses on the curved spacetime generalization Derrick-type (scaling) virial identities, and their relations to generalizations of~\eqref{Deser}, bringing, in this way, a different perspective on this type of GR virial identities. 

\medskip

This paper is organized as follows. In Section~\ref{S2} we start by reviewing the analysis in~\cite{Herdeiro:2021teo} on obtaining  generic virial identity from  Derrick-type scaling arguments in spherical symmetry, using the \textit{effective action} (EA) perspective followed in~\cite{Herdeiro:2021teo}. This leads to a generic form of the virial identity - $cf.$ eq.~\eqref{virialea4}, valid for any matter model. A concrete illustration (electrovacuum) is given. Then, we derive the Deser-type energy-momentum balance identity~\eqref{tmnid1} and propose that, for any matter model it coincides with the scaling virial identity.  In Section~\ref{S3} we shall consider a similar analysis beyond spherical symmetry, first using the EA framework, deriving a generic form for the scaling virial identity. Applying this to a specific example (flat spacetime $Q$-balls) illustrates its equivalence to~\eqref{Deser}. Then, we consider gravitating configurations in a particular metric form and derive the scaling virial identity for an illustrative case (BHs with synchronised scalar hair~\cite{Herdeiro:2014goa,Herdeiro:2015gia}). We are then able to show that the full scaling identity is a linear combination of the Einstein equations. This is then proposed as the master form for the virial identity within this gauge. We close in Section~\ref{S4} with a discussion of our results.

\medskip

 Throughout this paper we use units with $G=1=c$.

\section{Spherical symmetry}\label{S2}

Focusing on GR minimally coupled to matter we shall consider the following action 
\begin{eqnarray}  
\label{action} 
\mathcal{S}= \frac{1}{16 \pi} \int_{\mathcal{M}} d^4x \sqrt{-g}R  +\mathcal{S}_{\rm matter}[g,\psi] \ ,
\end{eqnarray}
which is the sum of an Einstein-Hilbert (EH) term
and a matter contribution (with $\psi$ collectively denoting matter fields), where $\mathcal{M}$ is the spacetime manifold.
The action (\ref{action})
should
be augmented to include the Gibbons-Hawking-York (GHY) 
boundary term~\cite{york1972role,Gibbons:1976ue,gibbons1993action}. 
The latter, however, does not contribute to the field equations, obtained from varying~\eqref{action}, which include the Einstein equations
\begin{eqnarray}  
\label{Einstein-eqs}
E_\mu^\nu\equiv G_\mu^\nu-8 \pi T_{\mu}^\nu=0 \ , 
\end{eqnarray}
where the Einstein tensor is $G_\mu^\nu\equiv R_{\mu}^\nu-\frac{1}{2}R \delta_{\mu}^\nu$, 
together with the matter field equations.

\subsection{The generic virial identity from the Derrick-type scaling argument}\label{S21}

In a previous paper~\cite{Herdeiro:2021teo}, we have presented an introduction to virial identities in field theory. Such identities serve a variety of purposes, including establishing no-go theorems for both solitonic and BH solutions, as well as serving as guides to finding new solutions of the equations of motion and checking the accuracy of numerical solutions.

The discussion in~\cite{Herdeiro:2021teo} found itself a natural arena in the context of \textit{effective actions} (EAs). 
As a brief summary of the results therein consider a one-dimensional (1D) EA, obtained after plugging an ansatz with spherical symmetry in~\eqref{action},  of the form
	\begin{equation}
	 \mathcal{S}^{\rm eff}[q_j(r),q'_j(r),r] = \int_{r_i}^{\infty}\hat{\mathcal{L}}\left( q_j,q'_j,r\right)dr \ ,
\label{actionspatial3}
	\end{equation}
where $q_j$ are a set of ${\cal N}$ parameterizing functions of some configuration, $j=1\dots {\cal N}$,  
$i.e.$ the radial metric functions and the ones in the matter ansatz. They depend on a single "radial" coordinate $r$ and $r_i$ is some appropriately chosen constant\footnote{In the examples considered in~\cite{Herdeiro:2021teo}, $r_i=0$ for solitons and $r_i=r_H$ for BHs, where $r_H$ is the event horizon radial coordinate.}. We use the notation $q_j'\equiv dq_j/dr$, and the \textit{effective Lagrangian} $\hat{\mathcal{L}}$ contains a total derivative term
\begin{equation}
\hat{\mathcal{L}}\left(q_i,q'_i,r\right)=\mathcal{L}\left(q_i,q'_i,r\right)+\frac{d}{dr}f\left(q_i,q'_i,r\right) \ ,
\label{lplusf}
\end{equation}
with $f$ being some function that depends on the same variables as the non-total derivative piece of the effective Lagrangian, $\mathcal{L}$. Then, a scaling transformation 
	\begin{equation}
	 r\rightarrow \tilde{r} = r_i + \lambda(r-r_i) \ ,
\label{scaling1}
	\end{equation}
induces a variation of any fiducial configuration $q_j(r)$, as $q_j(r)\rightarrow q_{\lambda j}(r)=q_j( \tilde{r})$.
The EA of the scaled configuration $q_{\lambda j}(r)$ becomes a \textit{function} of $\lambda$, denoted as $\mathcal{S}_\lambda^{\rm eff}$, and the true profile obeys the \textit{stationarity condition}
	\begin{equation}
	 \frac{\partial \mathcal{S}^{\rm eff}_\lambda}{\partial \lambda} \bigg|_{\lambda=1}=0 \ ,
\label{variationscale}
	\end{equation}
which yields the virial identity
\begin{equation}
\int_{r_i}^\infty\left[ \sum_j \frac{\partial \mathcal{L}}{\partial q'_j} q'_j -\mathcal{L} -\frac{\partial \mathcal{L}}{\partial r}(r-r_i)\right]dr = \left[\frac{\partial f}{\partial r}(r-r_i) - \sum_i\frac{\partial f}{\partial q'_i} q'_i\right]^{+\infty}_{r_i} \ . 
\label{virialea4}
\end{equation}

For spherically symmetric configurations, eq.~\eqref{virialea4} can be readily applied to field theory models yielding their virial identity. In the case of GR, the fact that the EH action contains second derivatives of the metric, implies that the total derivative term in~\eqref{lplusf}, defined  by $f$, is non-zero. Then, in general, one must consider the GHY term as part of the gravitational action, to obtain the correct virial identity. As shown in~\cite{Herdeiro:2021teo}, however, for specific gauge choices and under appropriate boundary conditions, the GHY term does not contribute. This is precisely the case for the gauge choice~\eqref{metric-Schw} and the boundary conditions we shall be interested in.

\subsubsection{Example: electrovacuum}\label{S211}
As a concrete example, that we shall recover in the next subsection, consider the matter Lagrangian in~\eqref{action} to be that of Maxwell's theory:
\begin{equation}
\mathcal{S}_{\rm matter}[g,A]=-\frac{1}{16\pi}\int d^4 x \sqrt{-g} F_{\mu\nu}F^{\mu\nu} \ ,
\label{maxaction}
\end{equation}
where $F=dA$ is the Maxwell field strength 2-form. Under spherical symmetry we take the ansatz~\eqref{metric-Schw}
and 
\begin{equation}
A=V(r)dt \ .
\label{epot}
\end{equation}
If this describes an asymptotically flat BH spacetime then there is a radial coordinate $r=r_H>0$, such that at the horizon ($r=r_H$),
\begin{equation}
N(r_H)=0 \ , \qquad  {\rm and} \qquad \sigma (r_H)\neq 0 \ ,
\label{bc1}
\end{equation}
and asymptotically ($r\rightarrow \infty$), 
\begin{equation}
m(r)\simeq M+\mathcal{O}\left(\frac{1}{r}\right) \ , \qquad \sigma(r) \simeq 1+\mathcal{O}\left(\frac{1}{r^2}\right) \ .
\label{bc2}
\end{equation}

Following~\cite{Herdeiro:2021teo} one can obtain the effective Lagrangian, apply~\eqref{virialea4} (considering the contribution of the GHY boundary term) which yields the virial identity:
\begin{equation}
\int_{r_H}^\infty \frac{r(V')^2}{\sigma}(2r_H-r) dr=0 \ ,
\label{virialev}
\end{equation}
where $V'\equiv dV/dr$ and we have already used the boundary conditions to eliminate a boundary term. One can check that for the Reissner-Nordstr\"om solution the integrand in~\eqref{virialev} is non-zero but the integral vanishes, obeying the identity. Thus, one can face~\eqref{virialev} as the virial identify for~\eqref{action} with~\eqref{maxaction}, under the boundary conditions~\eqref{bc1}-\eqref{bc2}.

\subsection{The generic virial identity as a Deser-type energy-momentum balance}\label{S23}
We shall now propose that identities such as~\eqref{virialev} are actually non-trivial re-arrangements of trivial~\textit{local} identities. In the process we obtain a generalization of Deser's identity~\eqref{Deser}.

Our starting point is the (non-trivial) observation that a certain combination of the Einstein tensor components
 $G_\mu^\nu$, 
for the metric gauge~\eqref{metric-Schw}, yields a total derivative:
\begin{equation}
r^2 \sigma
\left\{
\bigg(1-\frac{r_H}{r}\bigg)
\left[
2(G_t^t-G_\mu^\mu)+\left(\frac{1}{N}-1\right)(G_r^r-G_t^t)
\right]
-\frac{2r_H}{r}G_r^r
\right\}
=\frac{d}{dr}
\Big[
4(r-r_H) (\sigma m'-N r \sigma')
\Big] \ .
\label{georel}
\end{equation}
Here $r_H$ could be any constant. But we choose this constant to be the horizon radius, at which the boundary conditions~\eqref{bc1}-\eqref{bc2} are obeyed. Then, it holds that the radial integral of this total derivative from the horizon to infinity vanishes:
\begin{equation}
\int_{r_H}^\infty \frac{d}{dr}
\Big[
4(r-r_H) (\sigma m'-N r \sigma')
\Big] = \Big[
4(r-r_H) (\sigma m'-N r \sigma')
\Big]\big|_{r_H}^\infty \stackrel{\eqref{bc1}-\eqref{bc2}}{=} 0 \ .
\label{georel2}
\end{equation}
The vanishing of the integral of this total derivative still holds for regular self-gravitating solitons, for which $r_H=0$.

This preliminary observation leads us to consider the following (trivial) local identity 
\begin{eqnarray}
\label{s1n}
\bigg(1-\frac{r_H}{r}\bigg)
\left[
2(E_t^t-E_\mu^\mu)+\left(\frac{1}{N}-1\right)(E_r^r-E_t^t)
\right]
-\frac{2r_H}{r}E_r^r = 0 \ .
\end{eqnarray}
This is merely a combination of some of Einstein's equations~\eqref{Einstein-eqs}. Integrating this identity over a spacelike slice ($t=$constant), covering the exterior BH region, splitting the Einstein equations into the Einstein tensor and the energy-momentum tensor we obtain another identity:
\begin{eqnarray}
\label{iint}
\int_{r_H}^\infty
r^2 \sigma
\left\{
\left(1-\frac{r_H}{r}\right)
\left[
2(G_t^t-G_\mu^\mu)+\left(\frac{1}{N}-1\right)(G_r^r-G_t^t)
\right]
-\frac{2r_H}{r}G_r^r
\right\}dr
=
\\
\nonumber
8\pi 
\int_{r_H}^\infty
r^2 \sigma
\left\{
\left(1-\frac{r_H}{r}\right)
\left[
2(T_t^t-T_\mu^\mu)+\left(\frac{1}{N}-1\right)(T_r^r-T_t^t)
\right]
-\frac{2r_H}{r}T_r^r
\right\}dr \ .
\end{eqnarray}
In this identity, the integrand on either side is not necessarily zero. 
Then, making use of \eqref{georel}-\eqref{georel2} in \eqref{iint}, the left hand side of  \eqref{iint} vanishes and we are left with the identity~\eqref{tmnid1}, that involves solely the energy-momentum tensor (and metric elements).

\medskip

The non-trivial identity~\eqref{tmnid1} must be obeyed by \textit{any matter model} of the theory~\eqref{action}, described by the geometric form~\eqref{metric-Schw} and boundary conditions~\eqref{bc1}-\eqref{bc2}. For the particular case of the matter model~\eqref{maxaction}, with the ansatz~\eqref{epot}, one can easily see the non-trivial components of the energy-momentum tensor are:
\begin{eqnarray}
T_t^t=T_r^r=-T_\theta^\theta=-T_\varphi^\varphi\propto -\frac{V'^2}{2\sigma^2} \ .
\end{eqnarray}
It follows that~\eqref{tmnid1} becomes precisely the virial identity~\eqref{virialev}.

As another example, if instead of the Maxwell matter model one considers a real scalar field $\Phi$ theory under a potential $U(\Phi)$, with action
\begin{equation}
S_{\rm matter}=-\frac{1}{4\pi}\int d^4x \sqrt{-g}\Big[\partial_\mu \Phi \partial^\mu \Phi +U(\Phi)\Big] \ ,
\label{actionrs}
\end{equation}
then the non-trivial components of the energy momentum tensor are:
\begin{eqnarray}
T_t^t=T_\theta^\theta=T_\varphi^\varphi =-\frac{1}{8\pi}\left(N(\Phi')^2+U \right) \ , \qquad T_r^r =\frac{1}{8\pi}\left(N(\Phi')^2-U \right) \ ,
\end{eqnarray}
and~\eqref{tmnid1} becomes
\begin{equation}
\int_{r_H}^\infty r^2\sigma\left\{  (\Phi')^2\left[ 1-\frac{2r_H}{r}\left(1-\frac{m}{r}\right)  \right] +U\left[3-\frac{2r_H}{r}\right]\right\}dr=0 \ .
\label{nohair}
\end{equation}
This is the virial identity first obtained via a scaling argument in~\cite{heusler1992scaling,heusler1996no} (see also~\cite{Herdeiro:2015waa}), to establish a no-scalar hair theorem for BHs in gravity minimally coupled to the matter model~\eqref{actionrs}. As we can see, it coincides with~\eqref{tmnid1}.  In general, we have checked that~\eqref{tmnid1} coincides with the virial identity derived in Section 6.2 in 
\cite{Herdeiro:2021teo} for several BH models.

If instead of a BH we consider self-gravitating spherical solitons, then the identity~\eqref{tmnid1} simplifies to:
\begin{eqnarray}
\int_{0}^\infty
r^2 \sigma
\left[
2\big(T_t^t-T_\mu^\mu\big)+\left(\frac{1}{N}-1\right)\big(T_r^r-T_t^t\big)
\right]dr=0 \ ,
\label{tmnid2}
\end{eqnarray}
where we are now requiring the soliton to be regular at the origin so that $
r(\sigma m'-N r \sigma')$ vanishes at $r=0$, besides asymptotically flat, which yields the same conditions as before, eq.~\eqref{bc2}. We have checked that~\eqref{tmnid2}
coincides with the virial identity derived in Section 6.1 in 
\cite{Herdeiro:2021teo} for several solitonic models, 
including spherical bosonic (both for scalar and vector/Proca) stars as well as Dirac stars, 
by using Derrick's scaling procedure.

Finally, observe that for flat spacetime ($r_H=0$, $N=\sigma=1$),~\eqref{tmnid2} reduces to
\begin{eqnarray}
\int_{0}^\infty
r^2 
\big(T_t^t-T_\mu^\mu\big)
dr=0 \ ,
\label{Deser2}
\end{eqnarray}
which coincides with~\eqref{Deser} in spherical coordinates. Consequently,~\eqref{tmnid1} and \eqref{tmnid2} are curved spacetime generalizations of Deser's identity.

The examples given in this Section substantiate the following proposal: the virial identity obtained from Derrick's scaling argument for  $any$ BH or horizonless configuration in a spherically symmetric model written in the coordinates~\eqref{metric-Schw} and with appropriate  boundary conditions (regularity at the origin or horizon and asymptotically flat) is of the form~\eqref{tmnid1} (simplified to~\eqref{tmnid2} for solitons). This is therefore a \textit{master form for the virial identity, where it becomes clear that it is an energy-momentum balance condition}, generalizing Deser's identity~\eqref{Deser}.

\medskip

A natural question is then whether this lesson --  that virial identities obtained from Derrick's scaling procedure are equivalent to the generalized Deser's identity~\eqref{tmnid1} -- can be generalized beyond spherical symmetry. In the next section we will consider the case of stationary and axi-symmetric configurations to tackle this question.

%
%
\section{Beyond spherical symmetry}
\label{S3}
%

\subsection{The generic virial identity from the Derrick-type scaling argument}
\label{Section31}

Beyond spherical symmetry, for instance in axial-symmetry, one encounters higher dimensional EA. Thus, for generality, consider now an $n$-dimensional EA
	\begin{equation}
	 \mathcal{S}^{\rm eff}[q_j(r,\theta_\alpha);q'_j(r,\theta_\alpha),\partial_\alpha{q}_j(r,\theta_\alpha);r,\theta_\alpha] 
= \int\dots\int \prod_{\alpha=1}^{n-1} d\theta_\alpha \int_{r_i}^{\infty}\hat{\mathcal{L}}(q_j;q'_j,\partial_\alpha{q}_j;r,\theta_\alpha)dr \ ,
\label{eand}
	\end{equation}
	where $q_j$ ($j=1\dots {\cal N}$) are a set of ${\cal N}$ functions,  
	$\{\theta_\alpha\}$ are a set of $n-1$ variables on which the EA depends and the derivatives are denoted ${q_j'}(r,\theta_\alpha)\equiv \partial_r q_j(r,\theta_\alpha)$ and $\partial_\alpha{q_j}(r,\theta_\alpha)\equiv \partial_{\theta_\alpha} q_j(r,\theta_\alpha)$.
The effective Lagrangian is allowed to include a total \textit{radial} derivative 
\begin{equation}
\hat{\mathcal{L}}(q_j;q'_j,\partial_\alpha{q}_j;r,\theta_\alpha)={\mathcal{L}}(q_j;q'_j,\partial_\alpha{q}_j;r,\theta_\alpha)+\frac{d}{dr}f(q_j;q'_j,\partial_\alpha{q}_j;r,\theta_\alpha) \ .
\label{lplusf2}
\end{equation}

For obtaining a virial identity we have now more freedom in the variables we may scale. Here, we shall focus on the simplest case in which a single scaling in a single coordinate (the ``radial'' coordinate) is performed. 
Consequently, the methodology closely mimics that of  the spherical case.
In this spirit, we consider the scaling transformation~\eqref{scaling1}
which varies a fiducial configuration $q_j(r,\theta_\alpha)$ into  $q_{\lambda j}(r,\theta_\alpha)=q_j(r_i+\lambda(r-r_i),\theta_\alpha)$.  Following the discussion of Section 2 in~\cite{Herdeiro:2021teo}, we find, from the stationarity condition~\eqref{variationscale}, a virial identity 
	\begin{align}
 \int\dots\int \prod_{\alpha=1}^{n-1} d\theta_\alpha  \left\{\int_{r_i}^\infty\left[ \frac{\partial \mathcal{L}}{\partial r}(r-r_i) -\sum_i \frac{\partial \mathcal{L}}{\partial q'_i} q'_i +\mathcal{L}\right]dr -  \left[\frac{\partial f}{\partial r}(r-r_i) - \sum_i\frac{\partial f}{\partial q'_i} q'_i\right]^{+\infty}_{r_i} \right\}=0 \ . 
\label{virialnD}
	\end{align}
This is no more no less than the integral in the $\theta_\alpha$ coordinates of the virial identity for 1D EAs~\eqref{virialea4}. For the specific case of stationary and axially symmetric spacetimes, $n=2$ and the (only) $\theta$ integral in~\eqref{virialnD}  is, for the standard polar coordinate, between $0$ and $\pi$.\footnote{In this case, for ease of notation, we make $\theta_1\rightarrow \theta$.}

%
%
\subsubsection{Example: flat spacetime $Q$-balls and Derrick equals to Deser}

\label{Section311}
%
%
As a simple application of~\eqref{virialnD}  let us discuss a field theory in flat spacetime. As such, consider the $4$-dimensional Minkowski spacetime in standard spherical coordinates, for which the line element reads.
	\begin{equation}
	 ds^2 = -dt ^2 + dr^2 +r^2 \big( d\theta ^2 + \sin ^2 \theta d\varphi ^2 \big)\ . 
	\end{equation}

Consider a complex-scalar field model with a (yet unspecified) potential, described by the action
	\begin{equation}\label{complexaction}
	 \mathcal{S}_{\rm matter}^{\Phi^*} =\frac{1}{4\pi}
\int d^4x \sqrt{-g}  \bigg[-\frac{1}{2}g^{\mu\nu}(\partial_\mu\Phi\partial_\nu\Phi^* + 	\partial_\mu\Phi^*\partial_\nu\Phi)-U(|\Phi|)\bigg] \ .
	\end{equation}
Here $^*$ denotes complex conjugation.  Such a model allows spinning solitons known as spinning $Q$-balls~\cite{Volkov:2002aj, Kleihaus:2005me}, if one considers the scalar ansatz
	\begin{equation}\label{CPhiS}
	 \Phi (t,r,\theta,\varphi ) \equiv \phi (r,\theta ) e^{-i \omega t + i m \varphi }\ ,
	\end{equation}
	where  $\phi$ is the ($\theta$-dependent) scalar field amplitude, $\omega\in \mathbb{R}^+$ is the frequency defining the harmonic time-dependence and $m\in\mathbb{Z}$ is the azimuthal harmonic index.

Defining the EA as $\mathcal{S}_{\rm matter}^{\Phi^*} =-2\pi \int dt \mathcal{S}^{\rm eff}$,\footnote{The factor of $2\pi$ is arbitrary and chosen for convenience.} we obtain an EA of the type~\eqref{eand} with $r_i=0$,  $n=2$, $f=0$ and the effective Lagrangian 
\begin{equation}
\mathcal{L}(\phi; \phi',\hat{\phi}; r, \theta)=r^2\sin\theta\left[-\omega^2\phi^2+\phi_{,r}^2+\frac{{\phi}_{,\theta}^2}{r^2}+\frac{m^2\phi^2}{r^2\sin^2\theta}+U(|\phi |)\right] \ .
\end{equation}
 Then, applying~\eqref{virialnD} we obtain the virial identity
	\begin{equation}
	 \int _0 ^{\pi} d\theta\ \int _0 ^{+\infty} dr \, r^2 \sin \theta \Bigg[-3 \omega ^2 \phi ^2 +\phi_{,r}^2+\frac{{\phi}_{,\theta}^2}{r^2}+\frac{m^2\phi ^2}{r^2 \sin ^2 \theta }+ 3U(|\phi|)\Bigg]=0\ . 
\label{virialsqballs}
	\end{equation}
It can be easily checked that for spherical solutions, since ${\phi}_{,\theta}=0$ and $m=0$, then~\eqref{virialsqballs} reduces to the virial identity for $Q$-balls obtained in~\cite{Herdeiro:2021teo}. 

The identity~\eqref{virialsqballs} shows that even in axi-symmetry, the $\omega^2$ term is the only negative term for non-negative potentials. If the potential is just a mass term, $U(|\phi|)=\mu^2\phi^2$, the bound state condition $\omega^2<\mu^2$ implies that the integrand is everywhere positive, ruling out non-trivial solutions, just like in the static case. Thus, the existence of spinning $Q$-balls for non-negative potentials, also requires self-interactions such that $U(|\phi|)-\omega^2\phi^2$ becomes negative in some spacetime regions; rotation \textit{per se} cannot support such solitons without self-interactions.

Again, the virial identity~\eqref{virialsqballs} obtained from the scaling argument has another guise; it is actually equivalent to Deser's identity~\eqref{Deser} in axial symmetry. Indeed, the energy-momentum tensor obtained from the action~\eqref{complexaction} is
\begin{equation}
T_{\mu\nu}=\frac{1}{8\pi}\bigg\{\partial_\mu\Phi\partial_\nu\Phi^*+\partial_\nu\Phi\partial_\mu\Phi^*-g_{\mu\nu}\Big[\partial_\alpha\Phi\partial^\alpha\Phi^*+U(|\Phi|)\Big]\bigg\} \ .
\end{equation}
Then, a simple computation, with the ansatz~\eqref{CPhiS} shows that the virial identity~\eqref{virialsqballs} is equivalent to
\begin{equation}
\int dr d\theta  \, r^2\sin\theta \Big(T_r^r+T_\theta^\theta+T_\varphi^\varphi\Big) = 0 \ ,
\end{equation}
$i.e.$ Deser's identity~\eqref{Deser} in axial symmetry, as advertised.

\subsection{The virial identity from the Derrick-type scaling argument in a specific gauge}
\label{Section32}
%
The generic discussion in Section~\ref{Section31} did not have to specify the metric gauge. To make further progress and discuss GR in axial symmetry,
we shall consider generic equilibrium solitons and BHs that can be described by the stationary and axially symmetric ansatz: 
	\begin{equation}\label{AxmetricNum}
	 ds^2 = -e^{2F_0(r,\theta)}N(r) dt^2+ e^{2F_1(r,\theta)}\left[\frac{dr^2}{N(r)}+r^2d\theta^2\right] + e^{2F_2(r,\theta)}r^2\sin^2\theta\Big[d\varphi-W(r,\theta)dt\Big]^2 \ , \quad \;\;\; N(r)\equiv 1-\frac{r_H}{r} \ .
	\end{equation}
	
	Ansatz~\eqref{AxmetricNum} introduces  four parameterizing functions $F_i(r,\theta)$ and $W(r,\theta)$ and it 
	 has proven useful
	for computing numerical solutions. 
	Moreover, well known analytic solutions, such as the Kerr metric, can also be put in this form, as we shall recall below. Additionally, in this specific case the GHY and total derivative terms cancel out so they end up not contributing, as detailed in Section~\ref{Section34}. As such the boundary term will not be considered in the following discussion that will establish the master form of the virial identity.

\medskip

	Derrick's argument follows through the scaling of the radial component 
\begin{equation}
r\rightarrow r_\lambda= r_H+ \lambda(r-r_H) \ ,
\label{reescaling0}
\end{equation} and the metric/matter functions as
		\begin{align}\label{reescaling}
	 W_\lambda &(r,\theta) \rightarrow W(r_\lambda,\theta) \ ,\qquad 	F_{i\ \lambda} (r,\theta) \rightarrow F_i (r_\lambda,\theta)\ , \qquad N_\lambda(r) \rightarrow N(r_\lambda) \ ,\qquad 	 \psi_\lambda (r,\theta) \rightarrow \psi(r_\lambda,\theta)\ .
	\end{align}

To perform the scaling computation under ansatz~\eqref{AxmetricNum}, we observe that the following relation holds
\begin{eqnarray}  
\label{R-BH} 
 \sqrt{-g}R={\cal L}_{(g)}+\frac{\partial T_r}{\partial r}+ \frac{\partial T_\theta}{\partial \theta} \ ,
\end{eqnarray}
where we have introduced the gravity effective Lagrangian, 
\begin{eqnarray}  
\label{Lg-BH} 
{\cal L}_{(g)}\equiv  {\cal L}_{(g)}^F+{\cal L}_{(g)}^W \ ,
\end{eqnarray}
which is conveniently written as the sum of two terms
\begin{eqnarray}  
\nonumber
 {\cal L}_{(g)}^F&=&2 e^{F_0+F_2} \sin\theta
   \bigg\{
	N   \Big[
	r(F_{1,r}-F_{2,r})
	+r^2 (F_{1,r}F_{2,r}+F_{1,r}F_{0,r}+F_{0,r}F_{2,r})
	     \Big]
	+\frac{r_H}{2}(F_{1,r}-F_{0,r})
	\\
	&&
	F_{1,\theta}F_{2,\theta}+F_{1,\theta}F_{0,\theta}+F_{0,\theta}F_{2,\theta}
	+\cot \theta (F_{1,\theta}-F_{2,\theta})
	\bigg\} \ ,
	\\
	 {\cal L}_{(g)}^W&=& \frac{1}{2} e^{-F_0+3F_2} r^2 \sin^3\theta
	\left(
	r^2 W_{,r}^2+\frac{1}{N}W_{,\theta}^2
	\right) \ ,
\end{eqnarray}
and we have defined  
\begin{eqnarray}  
&&
\label{Tr}
T_r=-2 e^{F_0+F_2}r^2N \sin\theta (F_{0,r}+F_{1,r}+F_{2,r}) \ ,~~
\\
\nonumber
&&
T_\theta=-2 e^{F_0+F_2}  \sin\theta (F_{0,\theta}+F_{1,\theta}+F_{2,\theta}) \ .
\end{eqnarray}
Apart from the metric form~\eqref{AxmetricNum},
one needs to specify the action and an ansatz for the matter fields $\psi$.
This ansatz is  $not$ necessarily axially symmetric, $i.e.$
the matter fields
 may possess an explicit dependence
on $\varphi$ and $t$.
We impose, however, that the energy-momentum tensor 
and the effective Lagrangian for the matter field(s)  
${\cal L}_{(\psi)}$
depend on $(r,\theta)$, only.
 
Then the (essential) equations of the model can be obtained 
from a 2D effective action
\begin{eqnarray}  
\label{Seff}
S^{\rm eff} =S^{\rm eff}_{(g)}+S^{\rm eff}_{(\psi)}~,
\end{eqnarray}
where
\begin{eqnarray} 
\label{Seffg} 
S^{\rm eff}_{(g)}= \frac{1}{16\pi } \int d^3x
\left(
{\cal L}_{(g)}^F+ {\cal L}_{(g)}^W
\right)~,
\end{eqnarray}
and in order to simplify some relations we introduce the compact notation
\begin{eqnarray} 
\int d^3x= \int_{r_H}^\infty dr \int_0^\pi d \theta \int_0^{2\pi}d\varphi=2\pi \int_{r_H}^\infty dr \int_0^\pi d \theta  ~.
\end{eqnarray} 
The   Einstein equations for the functions  
 $\{F_i,W \}$
can be obtained from 
the effective action (\ref{Seff}) 
  (but not the supplementary
set of two constraint equations).
Then the standard Derrick procedure~\eqref{reescaling0}-\eqref{reescaling}
leads to the following virial identity
\begin{eqnarray}  
\label{virial-general-BH}
{\cal V}_{(g)}=16\pi {\cal V}_{(matter)} \ .  
\end{eqnarray}
Here, ${\cal V}_{(matter)}$ is the model dependent matter part; on the other hand, the geometric contribution, ${\cal V}_{(g)}$,
is universal, and reads
\begin{eqnarray}  
\label{virial-g-BH}
{\cal V}_{(g)}\equiv {\cal V}_{(g)}^F+3 {\cal V}_{(g)}^W \ ,
\end{eqnarray}
where 
\begin{eqnarray}  
{\cal V}_{(g)}^F
&=&
\int d^3 x
  ~e^{F_0+F_2} 2\sin\theta
\bigg[
 	N r \left\{
	 F_{1,r}-F_{2,r} 
	+ N r  \left( F_{1,r}F_{2,r}+F_{1,r}F_{0,r}+F_{0,r}F_{2,r}\right)
	     \right\}
	\nonumber \\
	&&
	{~~~~~~~~~~~~~~~~~~~~~~}
	+F_{1,\theta}F_{2,\theta}+F_{1,\theta}F_{0,\theta}+F_{0,\theta}F_{2,\theta}
	+\cot \theta (F_{1,\theta}-F_{2,\theta})
	\bigg] \ ,
\\
{\cal V}_{(g)}^W
&=&
 \frac{1}{2}\int d^3 x \,
 e^{-F_0+3F_2} r^2 \sin^3\theta
	\left[
	\left(N-\frac{r_H}{3r}\right)r^2 W_{,r}^2+ W_{,\theta}^2
	\right] \ .
\end{eqnarray}

Thus, the general virial identity obtained from Derrick's scaling argument, for any stationary, axi-symmetric geometry written in the form~\eqref{AxmetricNum} is~\eqref{virial-general-BH}.

 \subsubsection{Example: Kerr black holes with (or without) scalar hair}
As a specific example consider first the vacuum case. Then,  all GR vacuum BH solutions (regular on and outside an event horizon) belong to the Kerr family~\cite{Kerr:1963ud}. The virial identity~\eqref{virial-general-BH} simplifies to
\begin{eqnarray}
\label{virial-kerr}
{\cal V}_{(g)}=0 \ . 
\end{eqnarray}
One can then show, using the explicit form of the Kerr metric for the ansatz~\eqref{AxmetricNum} given in \cite{Herdeiro:2015gia}, that the  integral 
(\ref{virial-kerr})
vanishes, even though the integrands in both
${\cal V}_{(g)}^F$
and
$ {\cal V}_{(g)}^W$
are non-vanishing.

As a non-vacuum case where the Kerr BH can be embedded in a larger family of regular (on and outside an event horizon) BH solutions, we consider Kerr BHs with synchronised scalar hair~\cite{Herdeiro:2014goa,Herdeiro:2015gia}. The matter model consists of a complex scalar field, with a canonical kinetic term, minimally coupled to Einstein's gravity, described by the action~\eqref{complexaction}. The specific choice of the scalar field potential $U(|\Phi|)$ is of no importance for our discussion.
This model allows for hairy BHs bifurcating from the Kerr solution,
under the scalar field ansatz~\eqref{CPhiS}. Then, one finds the following expression for the scalar field's contribution 
 to the virial identity 
(\ref{virial-general-BH}):
\begin{eqnarray}  
\label{virialscalar}
{\cal V}_{(matter)}
\equiv
{\cal V}_{(scalar)}
=
\int d^3x~
 e^{F_0+F_2} \sin \theta
\bigg[
r^2N^2\phi_{,r}^2+\phi_{,\theta}^2
+
e^{2(F_1-F_2)}\frac{m^2 \phi^2}{\sin^2 \theta}
\\
\nonumber
{~~~~~~~~~~~~~~~~~~~~~~~~~~~~~}
+3e^{2F_1}r^2
                            \left\{ 
 \left(1-\frac{2r_H}{3r}\right) U(\phi^2)- e^{-2F_0}( w-mW)^2 \phi^2 
                            \right\}
\bigg]~.
\end{eqnarray}

In the flat spacetime limit, the virial identity (\ref{virial-general-BH}) reduces to solely the matter part
\begin{equation}
{\cal V}_{(scalar)}\Big|_{F_0=F_1=F_2=W=r_H=0}=0 \ . 
\end{equation} 
It is easy to check that this identity is exactly that obtained for spinning $Q$-balls~\eqref{virialsqballs}.

\subsection{A master virial identity in the chosen gauge}
\label{Section33}
The virial identity 
(\ref{virial-general-BH}) with~\eqref{virial-g-BH} and~\eqref{virialscalar} is long and its form is not particularly enlightening. 
Moreover, our treatment of the axially symmetric case has departed from the simple conclusion in the spherical case where we have noticed that the virial identity was an energy-momentum balance condition. Still, is there some underlying principle connecting the two cases, which could therefore be thought of as generic?

In fact, we have unveiled a common structure. Let us built it from the example of the previous subsection. We have observed that one can express the matter part of that virial identity discussed in the previous subsection, $i.e.$ \eqref{virialscalar}, as
\begin{eqnarray}  
{\cal V}_{(scalar)}=\int  d^3x 
\sqrt{-g} 
\left\{
\left(1-\frac{3r_H}{2r}\right)T_\varphi^tW
-\left[\left(1-\frac{r_H}{2r}\right)T_r^r+
N\left(T_\theta^\theta+T_\varphi^\varphi\right)+\frac{r_H}{2r}T_t^t\right]
\right\} \ .
\label{virialm}
\end{eqnarray}
This integral depending (linearly) on components of the energy momentum tensor and on other metric elements is very similar, in spirit, to the left hand side of the generalized Deser identity~\eqref{tmnid1}. In the case at hand, however, and unlike the spherically symmetric case, there is also a contribution from the geometric part to the complete virial identity. Remarkably, we could check that a similar form to~\eqref{virialm} holds (up to a constant factor) for the gravitational part~\eqref{virial-g-BH}, simply replacing in~\eqref{virialm} the energy-momentum tensor components $T_\mu^\nu$ by the corresponding Einstein tensor components $G_\mu^\nu$.
As such, the virial relation~(\ref{virial-general-BH}) for the scalar field matter model~\eqref{complexaction}
reduces  to  {\it the integral of a suitable combination of Einstein 
equations} (subject to a set of boundary conditions).
$E_r^r$,
$E_\theta^\theta$,
$E_\varphi^\varphi$
and
$E_\varphi^t$:
\begin{equation}  
\label{f2} 
\int  d^3x  \sqrt{-g} 
\left\{
\left(1-\frac{3r_H}{2r}\right)E_\varphi^tW
-\left[\left(1-\frac{r_H}{2r}\right)E_r^r+N(E_\theta^\theta+E_\varphi^\varphi)+\frac{r_H}{2r}E_t^t\right]
\right\}= 0 \ .   
\end{equation}
Let us emphasize that it is a trivial fact that~\eqref{f2} is an identity. In fact the integrand is zero, not only the integral
(although both the Einstein 
and energy-momentum tensors are non-zero).
 But it is a rather non-trivial fact that it is the identity that is obtained by Derrick's scaling argument applied to the specific model we have used to derive it.

\medskip

We can now discuss the common structure for the virial identities in the spherical and axial cases. The virial identity obtained from the scaling argument can be recast as a (trivial) identity which is a combination of the corresponding Einstein equations, $cf.$~\eqref{iint} and~\eqref{f2}. In the spherical case, however, the geometric part could be cast as a total derivative such that, under appropriate boundary conditions, it vanishes, leaving a non-trivial identity in terms of the energy-momentum tensor~\eqref{tmnid1}. In the axially symmetric case, and in the chosen gauge, this last step could not be done. 

\medskip

Furthermore, let us also emphasise that 
we have only demonstrated that~\eqref{f2} 
is the scaling virial identity for a specific model, 
with the matter action~\eqref{complexaction}. For a general matter model~\eqref{f2} remains obviously as an identity. We are \textit{conjecturing} that~\eqref{f2}  remains as the generic scaling virial relation for any asymptotically flat BH solution in GR that can be written in the form~\eqref{AxmetricNum} and with appropriate boundary conditions, regardless of the matter content.

\medskip

It is worth discussing the specific case of solitons. The model~\eqref{action} with the matter part given by~\eqref{complexaction} has self-gravitating soliton solutions known as boson stars. 
Restricting to an axially symmetric geometry, these solutions, both static and spinning, can be studied with the metric parametrization~\eqref{AxmetricNum} and $r_H=0$, $e.g.$~\cite{Herdeiro:2015gia,Herdeiro:2016tmi,Herdeiro:2019mbz}.

In this solitonic case, $r_H=0$ and the virial identity obtained from the scaling argument~\eqref{virial-general-BH} or, alternatively,~\eqref{f2},  simplifies to 
\begin{equation}  
\label{f3} 
\int  d^3x  \sqrt{-g} 
\left\{E_\varphi^tW+E_t^t
-E_\mu^\mu
\right\}= 0 \ .     
\end{equation}
Again, we \textit{conjecture} that this generic identity is the scaling virial relation for \textit{any} asymptotically flat solitonic solution in GR, regardless of the matter content, which can be written in the form~\eqref{AxmetricNum} with $r_H=0$.
 Apart from boson stars~\cite{schunck2003general},
we have verified this conjecture for  
Proca stars~\cite{brito2016proca}
and
Dirac stars~\cite{finster1999particlelike} 
(see also~\cite{herdeiro2017asymptotically,Herdeiro:2019mbz,Herdeiro:2020jzx}).
However, we expect it to hold for 
any other solitonic solutions arising in specific matter models.

The master identity~\eqref{f3} yields in the flat spacetime limit, 
$i.e.$ with
$F_i=W=0=r_H$ in~\eqref{AxmetricNum},
\begin{eqnarray}  
\label{virial-flat}
\int d^3 x\sqrt{-g}
\left(
T_t^t-T_\mu^\mu
\right) =0 \ .
\end{eqnarray}
In Cartesian coordinates this is relation~\eqref{Deser}. Thus, again, we see the Deser identity to be a special case of the master virial identity~\eqref{f2}. 

For completeness, let us summarize Deser's argument, which uses a different route from scaling arguments.
Working in Cartesian coordinates $x^i$ ($i=1,2,3$),
one assumes the existence of a stationary soliton in some field theory model.
Then,  the following (trivial) identity\footnote{Observe
that Deser's argument cannot be extended to a curved spacetime background.}
\begin{eqnarray}
\label{Deser1}
\frac{\partial }{\partial x^i}\left(x^j T_j^i \right)=T_i^i+x^j \frac{\partial   T_j^i }{\partial x^i}~,
\end{eqnarray}
together with its volume integral is considered.
The left hand side of such integral, which integrates the divergence of the dilatations current, vanishes from regularity and finite energy requirements. 
The second term on the right hand side of (\ref{Deser1})
vanishes from energy-momentum conservation (plus staticity)
and thus we are left with Deser's identity~\eqref{Deser}.   
It also follows that the total mass-energy of a  soliton in $3+1$ dimensions
is determined by the trace of the energy-momentum tensor.

 \subsection{The boundary term contribution}
\label{Section34}
In the discussion above concerning axially symmetric solutions 
we have neglected the GHY boundary term. 
This is justified for the reason we shall now detail. 

For spacetimes of the form~\eqref{AxmetricNum}, 
the spacetime boundary is taken at some constant $r$,
with $r\to \infty$ at the end of the computation.
The induced boundary metric reads
\begin{eqnarray}
d\sigma^2=\gamma_{ij}dx^i dx^j = e^{2F_1} r^2 d\theta^2 +e^{2F_2}r^2 \sin^2 \theta (d\varphi-W dt)^2- e^{2F_0}Ndt^2 \ ,
\qquad  N=1-\frac{r_H}{r} \ ,
\end{eqnarray}
with
$
\sqrt{-\gamma}=e^{F_0+F_1+F_2} \sqrt{N} r^2 \sin \theta.
$
It follows that the normal vector and the extrinsic curvature trace are:
 \begin{eqnarray}
n=e^{-F_1}\sqrt{N}\partial_{r} \ , \qquad
K=\frac{e^{-F_1}}{r\sqrt{N}} 
\left(
\frac{1}{2}rN'+2N+N r(F_{0,r}+F_{1,r}+F_{2,r})
\right) \ .
\end{eqnarray}

In the asymptotically flat case, the obvious  reference spacetime is Minkowski's.
That is, 
working in spherical coordinates, one takes the boundary to be 
	a sphere at radius $  r_0$,
	with
 \begin{eqnarray}
 d\sigma^2_{\rm (ref)}=-dt^2+  r_0^2 d\Omega_2^2 \ .
 \end{eqnarray}
The choice of $  r_0 $ is arbitrary. We take
$ r_0=r e^{F_1}$ and thus
 \begin{eqnarray}
K_0= \frac{2}{r }e^{-F_1} \ .
 \end{eqnarray}
It follows that
 \begin{eqnarray}
\label{KK0}
(K-K_0)\sqrt{-\gamma} =
e^{F_0+F_2}\sin \theta
\left[
2r N+
\frac{1}{2}r^2 N'-2r \sqrt{N}+N r^2(F_{0,r}+F_{1,r}+F_{2,r})
\right] \ .
 \end{eqnarray}

Following our earlier work~\cite{Herdeiro:2021teo},
this boundary term
should be included
in the bulk action,
and may provide a non-zero contribution  to the Derrick-type virial identity.
However, this is not the case for the 
employed
metric ansatz, 
once we observe that 
the gravity  action
also
contains a boundary term,
$cf.$ (\ref{R-BH}).
 As such,
 the effective gravity Lagrangian will 
contain a total derivative term
$df/dr$, with
 \begin{eqnarray}
f=T_r +2 (K-K_0)\sqrt{-\gamma},
 \end{eqnarray}
where
$T_r$
is 
given by (\ref{Tr}). 
A direct computation obtains
 \begin{eqnarray}
f= e^{F_0+F_2}\sin \theta 
\left[
4r (N-\sqrt{N})+r^2 N'
\right] \ .
 \end{eqnarray}
Recalling the general 
expression of the 
virial identity~\eqref{virialea4}
 (with $r_i=r_H$ for the case here)
 a simple computation shows 
that the right hand side term 
($i.e.$ the $f$-contribution in~\eqref{virialea4}) vanishes, 
both at the horizon and at infinity.

%
\section{Conclusions and discussion}
\label{S4}

In this work we have deconstructed virial identities obtained by scaling arguments in GR. Specifically, we have shown that when working in specific metric gauges, the virial identities obtained from Derrick-type scaling arguments can be recast as self-evident identities, since they are (non-trivial) combinations of the equations of motion. This has allowed us to obtain some master form identities -- eqs.~\eqref{tmnid1} and~\eqref{f2} -- that we propose are \textit{generic}, corresponding to the scaling virial identities for any matter model describing BHs or solitons with the appropriate boundary conditions in those metric gauges. 

\medskip

 In some cases, as in the case of spherical symmetry in the specific metric gauge~\eqref{metric-Schw}, the geometric part of the identity can be integrated out, resulting in a non-trivial energy-momentum balance condition~\eqref{tmnid1}, that can be thought of as a curved spacetime generalization of Deser's pressure balance condition for flat spacetime solitons~\eqref{Deser}.
This particular result seems to be associated with the special property of the metric form~\eqref{metric-Schw}, for which the EH action is invariant (up to a boundary term) under the scaling transformation~\cite{Herdeiro:2021teo}. 

\medskip

In the more generic case of axial symmetry, within the chosen metric gauge in this work ($cf.$ eq.\eqref{AxmetricNum}), it still holds that the virial scaling identity can be recast as a non-trivial combination of the Einstein equations ($cf.$ eq.~\eqref{f2}), but it is not true that the geometric part can be integrated out to yield a non-trivial energy-momentum balance condition. It will be interesting to understand whether a metric gauge exists (in axial symmetry) where it is possible to integrate out the geometric part. This stands out as an interesting direction for future research.

\medskip

Finally, it is worth emphasizing that this work unveils the true role of scaling virial identities in GR. Being (integrated) combinations of the equations of motion, they are, of course, not independent from the latter. Their usefulness in providing insight on the existence (or non-existence) of solutions is not challenged by this observation. The combinations of the equations of motion provided by the identity may collect useful arrangements of the unknowns to extract conclusions, as exemplified by $Q$-balls virial identity~\eqref{virialsqballs} and by the no-scalar hair virial identity~\eqref{nohair}. 
Moreover, they may also  provide an useful test for the
accuracy of numerical solutions. 

\section*{Acknowledgements}
This work is supported by the Center for Research and Development in Mathematics and Applications (CIDMA) through the Portuguese Foundation for Science and Technology (FCT - Fundaç\~{a}o para a Ci\^{e}ncia e a Tecnologia), references UIDB/04106/2020, UIDP/04106/2020.  We acknowledge support from the projects PTDC/FIS-OUT/28407/2017, CERN/FISPAR/0027/2019, PTDC/FIS-AST/3041/2020 and CERN/FIS-PAR/0024/2021. This work has further been supported by the European Union’s Horizon 2020 research and innovation (RISE) programme H2020-MSCA-RISE-2017 Grant No. FunFiCO-777740.

%
  \bibliographystyle{ieeetr}
  \bibliography{biblio}


\end{document}